\documentclass[12pt]{article}

\usepackage{cite}
\usepackage{epsfig}
\usepackage{amsmath}
\usepackage{lscape}

\newcommand{\LU}{\Lambda_{\cal U}}
\newcommand{\dU}{{d_{\cal U}}}

\def\mathswitch#1{\relax\ifmmode#1\else$#1$\fi}
\def\mathswitchr#1{\relax\ifmmode{\mathrm{#1}}\else$\mathrm{#1}$\fi}

\newcommand{\PZ}{\mathswitchr Z}

\newcommand{\Pe}{\mathswitchr e}

\newcommand{\MZ}{\mathswitch {M_\PZ}}

\newcommand{\me}{\mathswitch {m_\Pe}}

\newcommand{\SLASH}[2]{\makebox[#2ex][l]{$#1$}/}
\newcommand{\Dslash}{\SLASH{D}{.5}\,}

\newcommand{\lesim}{\,\raisebox{-.1ex}{$_{\textstyle <}\atop^{\textstyle\sim}$}\,}

\newcommand{\mycaption}[1]{\caption{\sl #1}}

\begin{document}
\thispagestyle{empty}

\def\thefootnote{\fnsymbol{footnote}}

\begin{flushright}
ZU-TH 20/07
\end{flushright}

\vspace{1cm}

\begin{center}

{\Large\sc {\bf Astro Unparticle Physics}}
\\[3.5em]
{\large\sc 
A.~Freitas
and
D.~Wyler
}

\vspace*{1cm}

{\sl
Institut f\"ur Theoretische Physik,
        Universit\"at Z\"urich, \\ Winterthurerstrasse 190, CH-8057
        Z\"urich, Switzerland
}

\end{center}

\vspace*{2.5cm}

\begin{abstract}

We investigate the effects of all flavor blind CP-conserving unparticle
operators on 5th force experiments, stellar cooling, supernova explosions and
compare the limits with each other and with those obtainable from collider
experiments. In general, astrophysical bounds are considerably stronger, however
they depend strongly on the dimension $\dU$ of the unparticle operator. While
for  $\dU = 1$, 5th force experiments yield exceedingly strong bounds, the bounds
from stellar and supernova cooling are more comparable for  $\dU = 2$, with
stellar cooling being most restrictive.  Bounds on vectorial unparticle
couplings are generally stronger than those on scalar ones.

\end{abstract}

\def\thefootnote{\arabic{footnote}}
\setcounter{page}{0}
\setcounter{footnote}{0}
\renewcommand\thesubsubsection{\arabic{subsubsection}}
\renewcommand{\section}{\subsubsection}

\newpage


\subsubsection{Introduction}

Recently the possible existence of a non-trivial scale-invariant sector with a
non-trivial fixed point was proposed by Georgi  \cite{Georgi:2007ek}. These new
fields, which couple weakly to Standard Model (SM) particles, are quite 
different from other extensions of the SM as they are not described in terms of
particles but rather by "unparticles".  A different, but in effect similar
deviation from the standard model has been proposed by Van der Bij 
\cite{vanderBij:2006ne}. The picture is valid up to a certain scale, above which the picture
changes. At low energies the unparticle sector is characterized by a scaling
dimension $\dU$, which is in general non-integer.

In this paper we want to assess possible effects of this extension of the
standard model in  astrophysics. There are by now several studies in this
direction \cite{Davoudiasl:2007jr, Hannestad:2007ys};  in this paper we combine
the different manifestations and give also a more detailed and complete
treatment of the various unparticle  operators. If the conformal invariance is
not broken in the infrared, as it is assumed throughout this paper,
astrophysical constraints can highly restrict the interactions between
unparticle and SM fields.

We consider only couplings between SM singlet unparticles and
Standard Model fields through CP-conserving and flavor blind interactions. In
Ref.~\cite{Chen:2007qr} a list of operators composed of SM fields with
dimensions 4 or less has been given. For our purpose we only need the couplings
to fermions and gauge bosons. With fermions we have:
\begin{align}
{\cal L}_{{\cal U}ff} &= \begin{aligned}[t]
\frac{{\cal C}_{\rm V}}{\LU^{\dU-1}} \,
 \bar{f} \gamma_\mu f \, O^\mu_{\cal U} +
\frac{{\cal C}_{\rm A}}{\LU^{\dU-1}} \,
 \bar{f} \gamma_\mu \gamma_5 f \, O^\mu_{\cal U} &+
\frac{{\cal C}_{\rm S1}}{\LU^{\dU}} \,
 \bar{f} \Dslash f \, O_{\cal U} +
\frac{{\cal C}_{\rm S2}}{\LU^{\dU}} \,
 \bar{f} \gamma_\mu f \, \partial^\mu O_{\cal U} \\
&+
\frac{{\cal C}_{\rm P1}}{\LU^{\dU}} \,
 \bar{f} \Dslash \gamma_5 f \, O_{\cal U} +
\frac{{\cal C}_{\rm P2}}{\LU^{\dU}} \,
 \bar{f} \gamma_\mu \gamma_5 f \, \partial^\mu O_{\cal U} \\
\end{aligned}
\\
& \equiv \begin{aligned}[t]
\frac{c_{\rm V}}{\MZ^{\dU-1}} \, \bar{f} \gamma_\mu f \, O^\mu_{\cal U} +
\frac{c_{\rm A}}{\MZ^{\dU-1}} \, \bar{f} \gamma_\mu \gamma_5 f \, O^\mu_{\cal U}
&+
\frac{c_{\rm S1}}{\MZ^{\dU}} \, \bar{f} \Dslash f \, O_{\cal U} +
\frac{c_{\rm S2}}{\MZ^{\dU}} \, \bar{f} \gamma_\mu f \, \partial^\mu O_{\cal U} 
\\ &+ 
\frac{c_{\rm P1}}{\MZ^{\dU}} \, \bar{f} \Dslash \gamma_5 f \, O_{\cal U} +
\frac{c_{\rm P2}}{\MZ^{\dU}} \, \bar{f} \gamma_\mu \gamma_5 f \, 
	\partial^\mu O_{\cal U}.
\label{eq:Uff}
\end{aligned}
\end{align}
Here the coefficients in eq.~\eqref{eq:Uff} have been scaled to a common mass, chosen as 
the $Z$-boson mass $\MZ$, so that the only
unknown quantities are the dimensionless coupling constants $c_{\rm X}$.
In the above equations, $D$ is the covariant derivative, which introduces
four-particle couplings involving the photon through $D_\mu = \partial_\mu + i
e Q A_\mu + \dots$
The term with $c_{\rm S2}$ does not contribute to physical
processes due to current conservation.

For photons we have
\begin{align}
{\cal L}_{{\cal U}\gamma\gamma} &= 
-\frac{{\cal C}_{\gamma\gamma}}{4\LU^{\dU}} \,
 F_{\mu\nu} F^{\mu\nu} \, O_{\cal U} 
-\frac{{\cal C}_{\tilde{\gamma}\tilde{\gamma}}}{4\LU^{\dU}} \,
 \epsilon^{\mu\nu\rho\sigma} F_{\mu\nu} F_{\rho\sigma} \, O_{\cal U} 
\\
&\equiv -\frac{c_{\gamma\gamma}}{4\MZ^{\dU}} \,
 F_{\mu\nu} F^{\mu\nu} \, O_{\cal U} 
-\frac{c_{\tilde{\gamma}\tilde{\gamma}}}{4\MZ^{\dU}} \,
 \epsilon^{\mu\nu\rho\sigma} F_{\mu\nu} F_{\rho\sigma} \, O_{\cal U} 
\end{align}
The term with $c_{\tilde{\gamma}\tilde{\gamma}}$ has the same structure as the
effective coupling of axions. Possible couplings to gluons are not considered.

In the following sections, we will consider the various standard tests of new
particles and forces and reach our conclusions. Since unparticle phase space
integration is more involved than for usual particles, we have added an appendix
with some useful results.


\subsubsection{Constraints from 5th force experiments}
\label{sc:5th}

New massless or light degrees of freedom can mediate new forces between SM
particles that lead to an effective modification of the Newtonian law of gravity
\cite{Okun:1969ey,Fischbach:1992fa}. The most stringent constraints come from
composition-dependent experiments, which were originally pioneered by
E\"otv\"os, Pek\'ar and Fekete \cite{Eotvos:1922pb}. They make use of the fact
that a new (fifth) force would in general act differently on different bodies of
equal mass, depending on their chemical composition \cite{Lee:1955vk}. Limits
for such an interaction have been derived on different length scales ranging
from sub-meter to astronomical scales of order AU.

For the experimental analyses, the fifth force is typically parametrized 
by the potential
\begin{equation}
V_5(r) = \pm \xi \, G \, m^2({}_1 H^1) \, B_i B_j \frac{e^{-r/\lambda}}{r},
\label{eq:yukpot}
\end{equation}
where the coupling constant $\xi$ has been normalized to the gravitational
interaction between two hydrogen ${}_1 H^1$ atoms, with $G$ Newton's
constant of gravity. $B_{i,j}$ are the baryon numbers of the two test objects.

The potential for interaction due to vector unparticle exchange can be derived
from its propagator \cite{Georgi:2007si,Cheung:2007ue}
\begin{align}
\Delta_{\rm F}^{\mu\nu} &\equiv \int {\rm d}^4 x \, e^{i p x} \langle
0 | T \, O^\mu_{\cal U}(x) O^\nu_{\cal U}(0) \rangle 
= i \frac{A_\dU}{2 \sin (\pi \dU)} \, 
\frac{-g^{\mu\nu} + p^\mu p^\nu / p^2}{(-p^2-i\epsilon)^{2-\dU}}, \\
A_\dU &= \frac{16\pi^{5/2}}{(2\pi)^{2\dU}} \, 
\frac{\Gamma(\dU+1/2)}{\Gamma(\dU-1)\Gamma(2\dU)}.
\end{align}
By taking the Fourier transform of this propagator in the low-energy limit one
obtains
\begin{align}
V_{\cal U} &= C \alpha_{\cal U} \, \frac{B_i B_j}{r^{2\dU-1}}, \label{eq:unppot}
\\
\alpha_{\cal U} &= \frac{c_{\rm V}^2}{4\pi} \MZ^{2-2\dU} \frac{A_\dU}{\pi} \,
\Gamma(2\dU-2) \\
&= \frac{\pi^{1/2}}{(2\pi)^{2\dU}} \, c_{\rm V}^2 \, \MZ^{2-2\dU}
 \frac{\Gamma(\dU-1/2)}{\Gamma(\dU)},
\nonumber
\end{align}
where $C={\cal O}(1)$ accounts for the quark and electron density inside the
nucleons and atoms of the test objects. For a conservative limit, we take $C \ge
1$.

Also $c_{\rm A}$ contributes in the same way as
$c_{\rm V}$, up to a prefactor, and yields
\begin{equation}
\alpha_{\cal U} = \frac{(3\pi)^{1/2}}{(2\pi)^{2\dU}} \, c_{\rm A}^2 \, \MZ^{2-2\dU}
 \frac{\Gamma(\dU-1/2)}{\Gamma(\dU)}.
\end{equation}
The scalar and pseudo-scalar interactions $\propto c_{\rm S2}, c_{\rm P1},
c_{\rm P2}$ do not contribute to long-range non-relativistic forces. However, 
the contribution from $c_{\rm S1}$ gives
\begin{equation}
\alpha_{\cal U} = \frac{\pi^{1/2}}{(2\pi)^{2\dU}} \, c_{\rm S1}^2 \, 
\frac{m_i m_j}{\MZ^{2\dU}}
 \frac{\Gamma(\dU-1/2)}{\Gamma(\dU)},
\end{equation}
where $m_{i,j}$ are the masses of the electrons and nucleons between which the
interaction is exchanged. The major contribution here comes from the nucleons
with $m_{i,j} \approx m({}_1 H^1) \approx \frac{1}{90} \MZ$.

The experimental limits on an interaction of type eq.~\eqref{eq:yukpot} can be
applied to the unparticle force eq.~\eqref{eq:unppot} by observing that the
constraints on eq.~\eqref{eq:yukpot} come mainly from measurements at a length
scale $r \approx \lambda$. For $r \gg \lambda$, $V_5$ is exponentially
suppressed, while for $r \ll \lambda$ the experiments are less sensitive
\cite{Fischbach:1992fa}. Therefore the exclusion limit at length scale $\lambda$
is
\begin{equation}
\alpha_{\cal U,\rm lim} \approx e^{-1} \, \xi_{\rm lim} \, G \, m^2({}_1 H^1) \;
\lambda^{2\dU-2}.
\end{equation}
This result agrees well with the power-law analysis in
Ref.~\cite{Adelberger:2006dh} for $\dU=2$.

\begin{figure}[tb]
\centering
\epsfig{figure=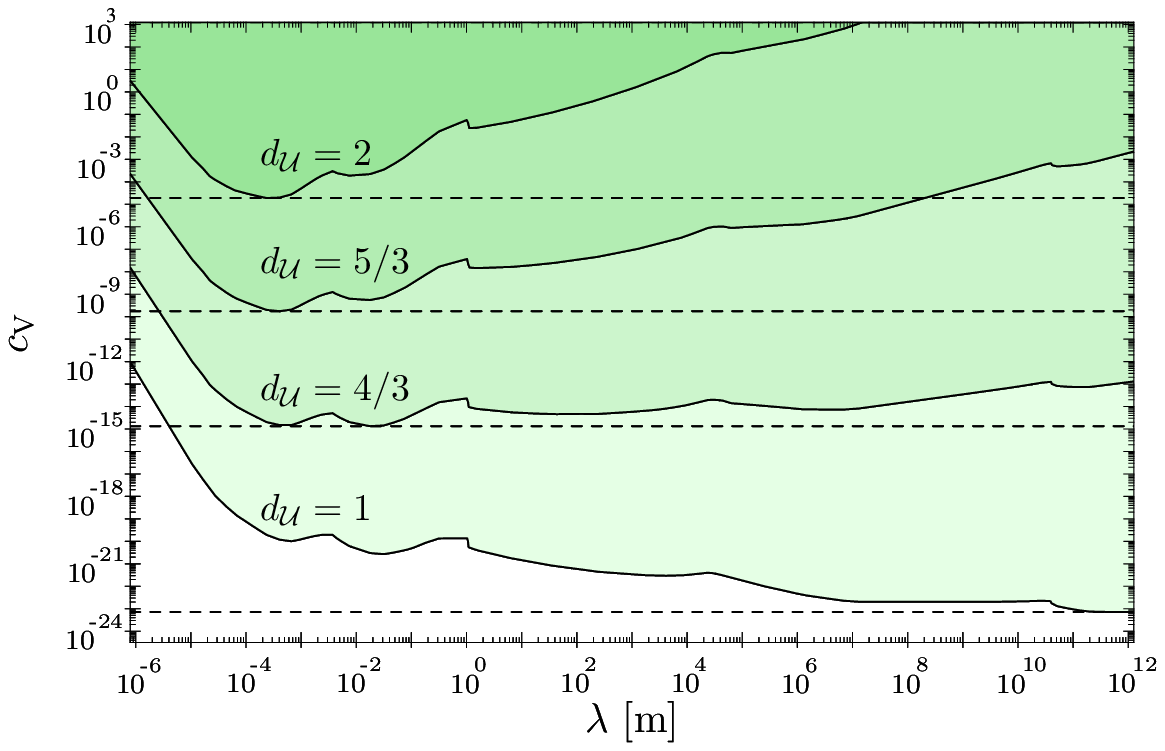, width=12cm, bb=20 470 360 680}
\mycaption{Limits on vector 
unparticle interactions from
E\"otv\"os-type fifth-force experiments at different length scales $\lambda$, 
for various scaling dimensions $\dU$.
The dashed lines indicate the overall limit derived from the whole $\lambda$
range.}
\label{fg:5th}
\end{figure}

Taking the experimental values (see Ref.~\cite{Fischbach:1992fa,Kapner:2006si} 
and references therein),
results are shown for different scaling dimensions in Fig.~\ref{fg:5th}.
They can be readily translated to the axial-vector and scalar cases.


\subsubsection{Constraints from stellar cooling}
\label{sc:stellar}

\paragraph{Constraints from stellar cooling on fermion couplings.}

In the hot and dense environment of stars, light weakly interacting particles
can be produced efficiently and would contribute to the cooling of the star.
Constraints on such particles can be derived from white dwarfs
\cite{Raffelt:1985nj,Isern:1992,Altherr:1993zd},  the ignition condition for
type I supernovae \cite{Wang:1992kn}, horizontal-branch stars with a
helium-burning core \cite{Raffelt:1990yz,Raffelt:1999tx,Raffeltbook}, and red
giants near helium ignition flash
\cite{Raffelt:1985nk,Dearborn:1985gp,Altherr:1993zd,Raffelt:1994ry}.  These
processes have been studied extensively for axion emission, with the strongest
bounds coming from helium-burning stars and red giants. In the following, we
will focus on the evaluation of unparticle emission from helium-burning stars,
which would lead to a reduction of the lifetime of the horizontal-branch stars.

\begin{figure}[tb]
\centering
\begin{tabular}{c@{\hspace{3em}}c@{\hspace{3em}}c} 
(a) & (b) & (c) \\
\raisebox{-2em}{\psfig{figure=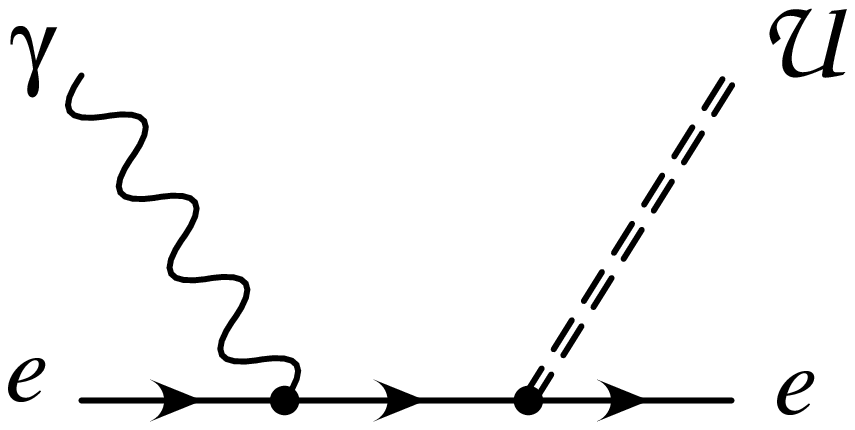, width=4cm}} & 
\psfig{figure=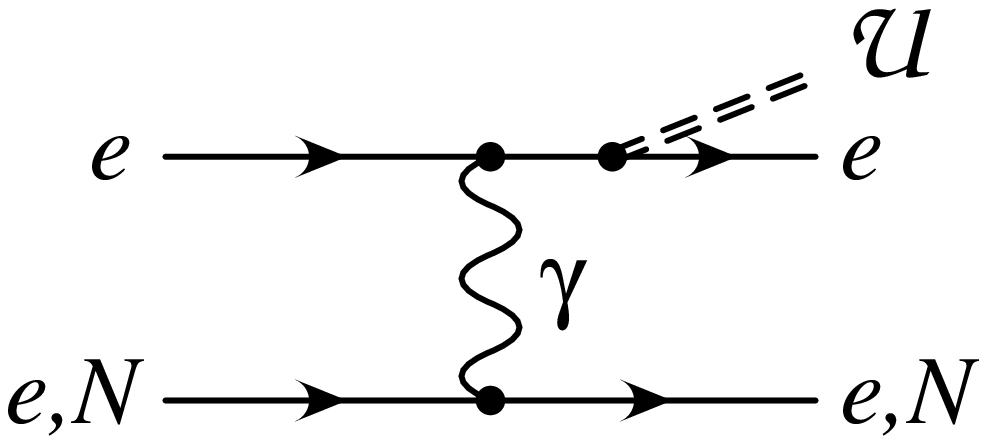, width=4.8cm} & 
\psfig{figure=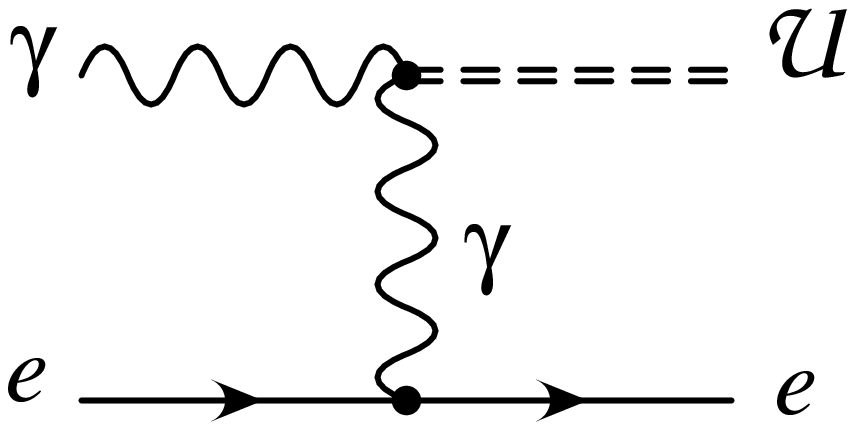, width=4cm} \\[-2em]
\raisebox{-2em}{\psfig{figure=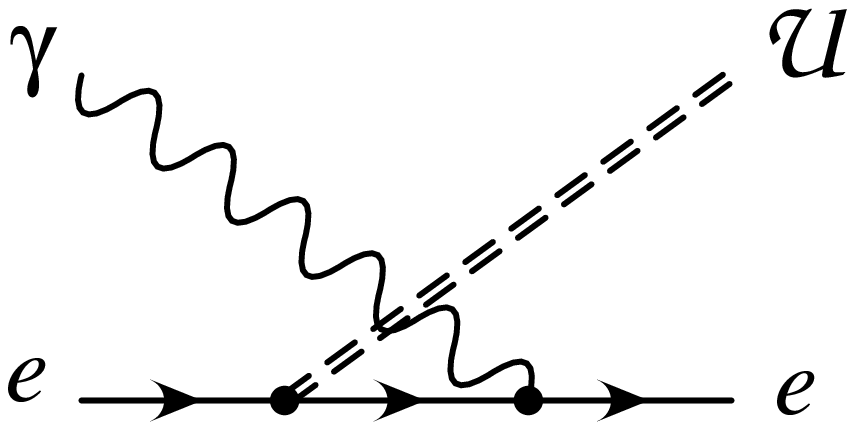, width=4cm}} & 
\psfig{figure=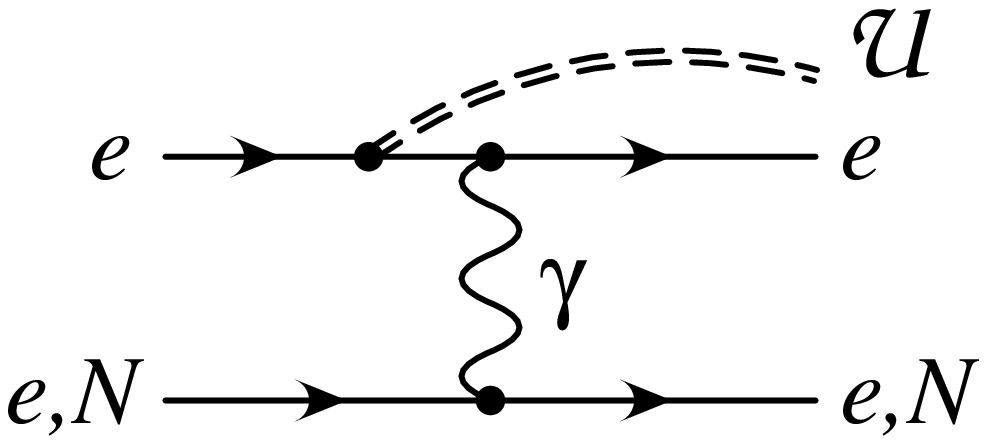, width=4.8cm} & 
\raisebox{0.5em}{\psfig{figure=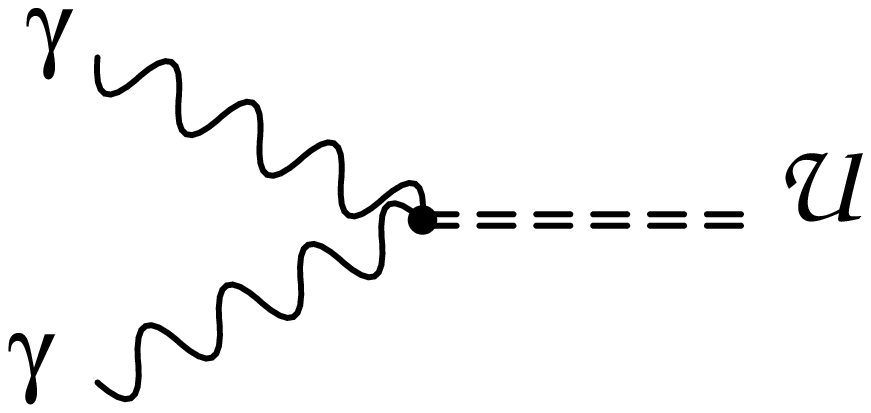, width=4cm}} \\[-2em]
\raisebox{-2em}{\psfig{figure=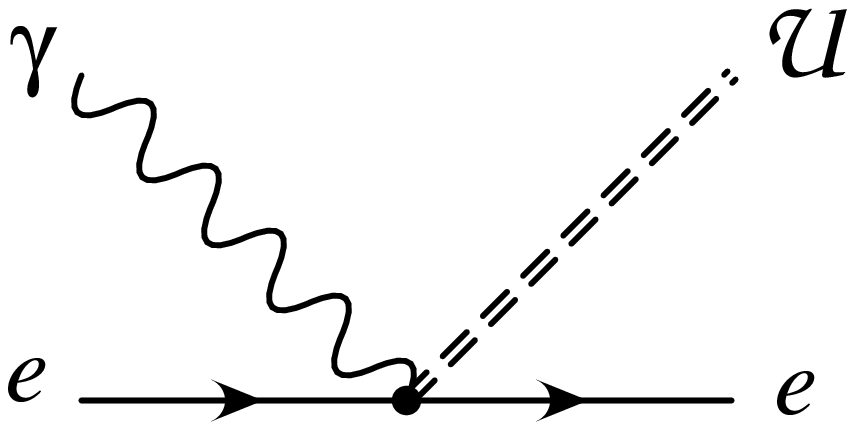, width=4cm}} & 
\psfig{figure=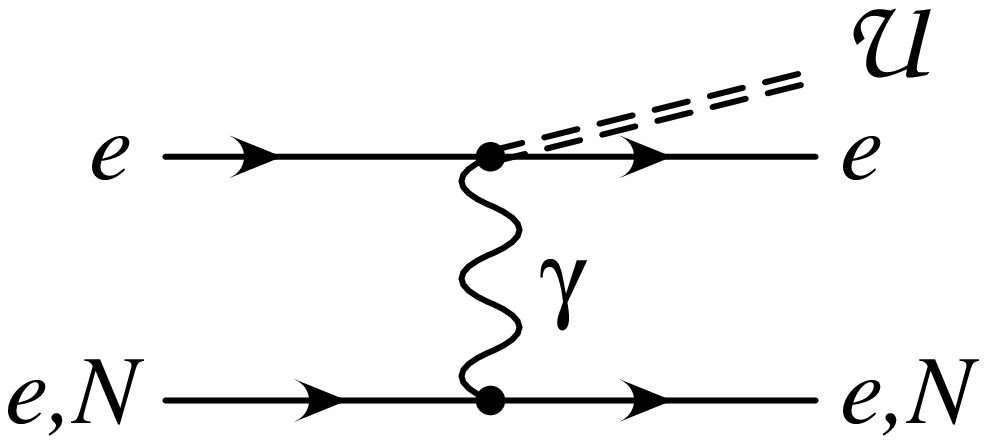, width=4.8cm} \\[-2.5em]
\end{tabular}
\mycaption{Feynman diagrams for unparticle emission in (a) Compton-like
processes, (b) bremsstrahlung-like processes and (c) processes with
unparticle-photon couplings.}
\label{fg:feyn1}
\end{figure}

Mainly two processes contribute to energy loss from horizontal-branch stars,
the Compton process $\gamma +e \to e + X$ and bremsstrahlung involving Hydrogen
and Helium nuclei as well as electrons, $e + {\rm H}^+ \to e + {\rm H}^+ + X$,
$e + {\rm He}^{2+} \to e + {\rm He}^{2+} + X$, $e + e \to e + e + X$. Here $X$
is the axion or unparticle. The corresponding Feynman diagrams are shown in
Fig.~\ref{fg:feyn1}~(a) and (b).

\vspace{1ex}
The total cross-section for axion emission through the Compton process is
\begin{equation}
\sigma^{\rm c}_{\rm a} = \frac{\alpha g^2_{aee}}{3\me^2} \, 
\left[\frac{\omega}{\me}\right]^2
\end{equation}
in the limit $\omega \ll \me$, where $\omega$ is the incoming photon energy.
$g_{aee}$ is the axion-electron coupling.

For unparticle emission, the calculations are somewhat more complicated. Due to
the phase space factor $A_{\dU} \theta(p^0_{\cal U}) \theta(p^2_{\cal U})
(p^2_{\cal U})^{\dU -2}$ \cite{Georgi:2007ek}, the final state integration
requires some care. The important integrals are collected in the appendix.
In the limit $\omega \ll \me$ one finds for the Compton
process production of unparticles
\begin{align}
\sigma^{\rm c}_{\rm {\cal U},V} &= \frac{\alpha c_{\rm V}^2}{\me^2} \, 
\frac{2\dU}{(1+2\dU)\Gamma(2\dU)} \,
\left[\frac{\omega}{2\pi\MZ}\right]^{2\dU-2}, \displaybreak[0]
\\
\sigma^{\rm c}_{\rm {\cal U},A} &= \frac{\alpha c_{\rm A}^2}{\me^2} \, 
\frac{2(2+\dU)}{(1+2\dU)\Gamma(2\dU)} \,
\left[\frac{\omega}{2\pi\MZ}\right]^{2\dU-2}, \displaybreak[0]
\\
\sigma^{\rm c}_{\rm {\cal U},S1} &= \frac{\alpha c_{\rm S1}^2}{\MZ^2} \, 
\frac{1}{(1+2\dU)\Gamma(2\dU)} \,
\left[\frac{\omega}{2\pi\MZ}\right]^{2\dU-2}, \displaybreak[0]
\\
\sigma^{\rm c}_{\rm {\cal U},P1} &= \frac{4\pi^2\alpha c_{\rm P1}^2}{\me^2} \, 
\frac{(2+2\dU+\dU^2)}{(1+2\dU)(3+2\dU)\Gamma(2\dU)} \,
\left[\frac{\omega}{2\pi\MZ}\right]^{2\dU}, \displaybreak[0]
\\
\sigma^{\rm c}_{\rm {\cal U},P2} &= \frac{16\pi^2\alpha c_{\rm P2}^2}{\me^2} \, 
\frac{(2+2\dU+\dU^2)}{(1+2\dU)(3+2\dU)\Gamma(2\dU)} \,
\left[\frac{\omega}{2\pi\MZ}\right]^{2\dU}, \displaybreak[0]
\end{align}
In the hot environment of a star photons are generated thermically, with a
distribution
\begin{equation}
n_\gamma(T,\omega) = \frac{\pi^{-2} \, \omega^2}{e^{\omega/T}-1}.
\end{equation}
The thermally averaged unparticle energy emission rate is then
\begin{equation}
Q(T)_{\rm c,\cal U} = \int_0^\infty {\rm d}\omega \, \omega \, n_e n_\gamma \,
\sigma^{\rm c}_{\cal U}(\omega),
\end{equation}
with the electron density
\begin{equation}
n_e \approx \frac{1+X_{\rm H}}{2} \, \frac{\rho}{m(\rm H)},
\end{equation}
where $\rho$ is the total density and $X_{\rm H}$ the mass fraction of
hydrogen.
The averaging gives
\begin{equation}
\sigma^{\rm c}_{\cal U}(\omega) = C \, \omega^r
\qquad
\Rightarrow
\qquad
Q^{\rm c}_{\cal U}(T) = C \, n_e \frac{\zeta(4+r) \,\Gamma(4+r)}{\pi^2} \, T^{4+r}.
\end{equation}

\vspace{1ex}\noindent
The emission rate for axion bremsstrahlung from electron-nucleus collisions
is, in the limit for small incident
electron velocities $\beta_i \ll \me$ \cite{Krauss:1984gm,Raffelt:1985nk}
\begin{equation}
Q^{\rm eZ}_{\rm a}(\beta_i) = \frac{2}{135\pi\me} \, Z^2\alpha^2 g^2_{aee}
n_e n_z \beta_i^5,
\end{equation}
where $n_{e,z}$ are the electron and nucleus densities and $Z$ is the proton
number of the nucleus.
For unparticle emission through bremsstrahlung in the non-relativistic limit
one finds
\begin{align}
Q^{\rm eZ}_{\rm {\cal U},V} &= \frac{Z^2\alpha^2 c_{\rm V}^2 \beta_i}{\me} \, 
n_e n_z \,
\frac{-8 (2+3\dU)\csc(2 \pi\dU)}{(2\dU-1)(1+2\dU)(3+2\dU)
\Gamma(2-2\dU)\Gamma(4\dU-1)} \,
\left[\frac{\me \beta_i^2}{\pi\MZ}\right]^{2\dU-2}, \displaybreak[0]
\\
Q^{\rm eZ}_{\rm {\cal U},A} &= \frac{Z^2\alpha^2 c_{\rm A}^2 \beta_i}{\me} \, 
n_e n_z \,
\frac{24 (1-\dU)\csc(2 \pi\dU)}{(2\dU-1)(1+2\dU)(3+2\dU)
\Gamma(2-2\dU)\Gamma(4\dU-1)} \,
\left[\frac{\me \beta_i^2}{\pi\MZ}\right]^{2\dU-2}, \displaybreak[0]
\label{eq:epseZA}
\\
Q^{\rm eZ}_{\rm {\cal U},S1} &= \frac{Z^2\alpha^2 c_{\rm S1}^2 \me \beta_i}{\MZ^2} \, 
n_e n_z \,
\frac{5 \pi^{-1/2}}{(2\dU-1)^2(1+2\dU)(3+2\dU)\Gamma(2\dU-1/2)} \,
\left[\frac{\me \beta_i^2}{\pi\MZ}\right]^{2\dU-2}, \displaybreak[0]
\\
Q^{\rm eZ}_{\rm {\cal U},P1} &= \frac{Z^2\alpha^2 c_{\rm P1}^2 \beta_i}{\me} \, 
n_e n_z \,
\frac{-\pi^2(15+14\dU+6\dU^2)\csc(2 \pi\dU)}{4(1+2\dU)(1+4\dU)
\Gamma(-2\dU)\Gamma(4\dU)\Gamma(\dU+7/2)} \,
\left[\frac{\me \beta_i^2}{\pi\MZ}\right]^{2\dU}, \displaybreak[0]
\\
Q^{\rm eZ}_{\rm {\cal U},P2} &= \frac{Z^2\alpha^2 c_{\rm P2}^2 \beta_i}{\me}
n_e n_z \,
\frac{-\pi^2(15+14\dU+6\dU^2)\csc(2 \pi\dU)}{(1+2\dU)(1+4\dU)
\Gamma(-2\dU)\Gamma(4\dU)\Gamma(\dU+7/2)} \,
\left[\frac{\me \beta_i^2}{\pi\MZ}\right]^{2\dU}. \displaybreak[0]
\end{align}
One can see that the rate for the axial vector vanishes for $d_U \to 1$; below
we will find this behaviour also for other processes. This results holds
however only for the leading power in the velocity $\beta_i$ in the
non-relativistic limit, as one may see when expanding the correct expression in
powers of $\beta$. The suppression can be understood from the fact that
electron-nucleon scattering is independent of the chirality of the particles
and therefore the $L-R$ coupling is suppressed. Because however unparticles for
$d_U > 1$ carry a third polarization degree of freedom, the suppression is not
total.

The bremsstrahlung emission rates have to be averaged over a Maxwellian
distribution\footnote{Since the density of horizontal-branch stars is relatively
low, screening and degeneracy (Pauli
blocking) effects are negligible \cite{Raffelt:1985nj,Raffelt:1994ry}.}
\begin{equation}
n_e(T,\beta_i) = \Bigl (\frac{m}{2 \pi T}\Bigr )^{3/2} 4\pi \beta_i^2 \,
\exp \Bigl (-\frac{m\beta_i^2}{2 T} \Bigr )
\end{equation}
so that
\begin{equation}
Q^{\rm eZ}_{\cal U}(\beta_i) = C \, \beta_i^r
\qquad
\Rightarrow
\qquad
\begin{aligned}[t]
Q^{\rm eZ}_{\cal U}(T) &= \int_0^\infty {\rm d}\beta_i \, n_e(T,\beta_i) \,
Q^{\rm eZ}_{\cal U}(\beta_i) \\ &=
C \, 2\pi^{-1/2} \, \Gamma(\tfrac{3+r}{2}) \, (2T/\me)^{r/2}.
\end{aligned}
\end{equation}
Furthermore, summing over the relevant nuclei,
\begin{equation}
n_e \, \sum_z Z^2 n_z \approx n_e (n_{\rm H} + 4 n_{\rm He})
 \approx \frac{1+X_{\rm H}}{2} \, \left(\frac{\rho}{m_{\rm H}}\right)^2.
\end{equation}
Bremsstrahlung in electron-electron collisions leads to very similar results as
bremsstrahlung in electron-nucleus collisions, except for the replacement $Z^2
n_e n_z \to 4 n_e^2$ in $Q(\beta_i)$ or
$Z^2 n_e n_z \to \sqrt{2} \, n_e^2$ in $Q(T)$, respectively 
\cite{Raffelt:1985nk}. Bremsstrahlung in nucleus-nucleus
collisions is negligible since the radiation of unparticles from nuclei with
mass $m_{\rm z}$ is suppressed by powers of $\beta_{i,\rm z}/\beta_{i,\rm e} \sim
(\me/m_{\rm z})^{1/2}$.

The impact of weakly interacting particle emission on star cooling can be
evaluated with a numerical code for stellar evolution
\cite{Dearborn:1985gp,Raffelt:1994ry}. For simplicity, we give here the
comparison of the unparticle emission rate to the axion emission constraints
which have been analyzed earlier
\cite{Raffelt:1990yz,Raffelt:1999tx,Raffeltbook}. The relation between the two
is summarized in Table~\ref{tab:redg}.

\renewcommand{\arraystretch}{1.2}
\begin{table}[tb]
\begin{center}
\begin{tabular}{|l||l|l|l|l|}
\hline
$\dU$ & 1 & 4/3 & 5/3 & 2  \\
\hline
$\frac{Q^{\rm c}_{\rm {\cal U},V}}{Q^{\rm c}_{\rm a}} \times 10^{12}$ &
$3.34 \, \frac{c_{\rm V}^2}{g^2_{aee}} \left(\frac{T}{\MZ}\right)^{-2}$ &
$1.69 \, \frac{c_{\rm V}^2}{g^2_{aee}} \left(\frac{T}{\MZ}\right)^{-4/3}$ &
$0.76 \, \frac{c_{\rm V}^2}{g^2_{aee}} \left(\frac{T}{\MZ}\right)^{-2/3}$ &
$0.32 \, \frac{c_{\rm V}^2}{g^2_{aee}}$ \\
$\frac{Q^{\rm c}_{\rm {\cal U},A}}{Q^{\rm c}_{\rm a}} \times 10^{12}$ &
$10.0 \, \frac{c_{\rm A}^2}{g^2_{aee}} \left(\frac{T}{\MZ}\right)^{-2}$ &
$4.22 \, \frac{c_{\rm A}^2}{g^2_{aee}} \left(\frac{T}{\MZ}\right)^{-4/3}$ &
$1.67 \, \frac{c_{\rm A}^2}{g^2_{aee}} \left(\frac{T}{\MZ}\right)^{-2/3}$ &
$0.64 \, \frac{c_{\rm A}^2}{g^2_{aee}}$ \\
$\frac{Q^{\rm c}_{\rm {\cal U},S1}}{Q^{\rm c}_{\rm a}} \times 10^{23}$ &
$5.25 \, \frac{c_{\rm S1}^2}{g^2_{aee}} \left(\frac{T}{\MZ}\right)^{-2}$ &
$1.99 \, \frac{c_{\rm S1}^2}{g^2_{aee}} \left(\frac{T}{\MZ}\right)^{-4/3}$ &
$0.72 \, \frac{c_{\rm S1}^2}{g^2_{aee}} \left(\frac{T}{\MZ}\right)^{-2/3}$ &
$0.25 \, \frac{c_{\rm S1}^2}{g^2_{aee}}$ \\
$\frac{Q^{\rm c}_{\rm {\cal U},P1}}{Q^{\rm c}_{\rm a}} \times 10^{12}$ &
$31.4 \, \frac{c_{\rm P1}^2}{g^2_{aee}}$ &
$18.4 \, \frac{c_{\rm P1}^2}{g^2_{aee}} \left(\frac{T}{\MZ}\right)^{2/3}$ &
$9.66 \, \frac{c_{\rm P1}^2}{g^2_{aee}} \left(\frac{T}{\MZ}\right)^{4/3}$ &
$4.71 \, \frac{c_{\rm P1}^2}{g^2_{aee}} \left(\frac{T}{\MZ}\right)^{2}$ \\
$\frac{Q^{\rm eZ}_{\rm {\cal U},P2}}{Q^{\rm eZ}_{\rm a}} \times 10^{12}$ &
$126 \, \frac{c_{\rm P2}^2}{g^2_{aee}}$ &
$73.5 \, \frac{c_{\rm P2}^2}{g^2_{aee}} \left(\frac{T}{\MZ}\right)^{2/3}$ &
$38.6 \, \frac{c_{\rm P2}^2}{g^2_{aee}} \left(\frac{T}{\MZ}\right)^{4/3}$ &
$18.8 \, \frac{c_{\rm P2}^2}{g^2_{aee}} \left(\frac{T}{\MZ}\right)^{2}$ \\
\hline
$\frac{Q^{\rm eZ}_{\rm {\cal U},V}}{Q^{\rm eZ}_{\rm a}} \times 10^{12}$ &
$118 \, \frac{c_{\rm V}^2}{g^2_{aee}} \left(\frac{T}{\MZ}\right)^{-2}$ &
$13.3 \, \frac{c_{\rm V}^2}{g^2_{aee}} \left(\frac{T}{\MZ}\right)^{-4/3}$ &
$2.04 \, \frac{c_{\rm V}^2}{g^2_{aee}} \left(\frac{T}{\MZ}\right)^{-2/3}$ &
$0.36 \, \frac{c_{\rm V}^2}{g^2_{aee}}$ \\
$\frac{Q^{\rm eZ}_{\rm {\cal U},A}}{Q^{\rm eZ}_{\rm a}} \times 10^{12}$ &
$0$ &
$2.21 \, \frac{c_{\rm A}^2}{g^2_{aee}} \left(\frac{T}{\MZ}\right)^{-4/3}$ &
$0.58 \, \frac{c_{\rm A}^2}{g^2_{aee}} \left(\frac{T}{\MZ}\right)^{-2/3}$ &
$0.14 \, \frac{c_{\rm A}^2}{g^2_{aee}}$ \\
$\frac{Q^{\rm eZ}_{\rm {\cal U},S1}}{Q^{\rm eZ}_{\rm a}} \times 10^{23}$ &
$185 \, \frac{c_{\rm S1}^2}{g^2_{aee}} \left(\frac{T}{\MZ}\right)^{-2}$ &
$17.4 \, \frac{c_{\rm S1}^2}{g^2_{aee}} \left(\frac{T}{\MZ}\right)^{-4/3}$ &
$2.29 \, \frac{c_{\rm S1}^2}{g^2_{aee}} \left(\frac{T}{\MZ}\right)^{-2/3}$ &
$0.36 \, \frac{c_{\rm S1}^2}{g^2_{aee}}$ \\
$\frac{Q^{\rm eZ}_{\rm {\cal U},P1}}{Q^{\rm eZ}_{\rm a}} \times 10^{12}$ &
$31.4 \, \frac{c_{\rm P1}^2}{g^2_{aee}}$ &
$9.22 \, \frac{c_{\rm P1}^2}{g^2_{aee}} \left(\frac{T}{\MZ}\right)^{2/3}$ &
$2.68 \, \frac{c_{\rm P1}^2}{g^2_{aee}} \left(\frac{T}{\MZ}\right)^{4/3}$ &
$0.77 \, \frac{c_{\rm P1}^2}{g^2_{aee}} \left(\frac{T}{\MZ}\right)^{2}$ \\
$\frac{Q^{\rm eZ}_{\rm {\cal U},P2}}{Q^{\rm eZ}_{\rm a}} \times 10^{12}$ &
$126 \, \frac{c_{\rm P2}^2}{g^2_{aee}}$ &
$36.9 \, \frac{c_{\rm P2}^2}{g^2_{aee}} \left(\frac{T}{\MZ}\right)^{2/3}$ &
$10.7 \, \frac{c_{\rm P2}^2}{g^2_{aee}} \left(\frac{T}{\MZ}\right)^{4/3}$ &
$3.09 \, \frac{c_{\rm P2}^2}{g^2_{aee}} \left(\frac{T}{\MZ}\right)^{2}$ \\
\hline
\end{tabular}
\end{center}
\mycaption{Comparison of unparticle emission rates to axion emission rates in a
stellar plasma of temperature $T$. Separately shown are the rates from
the Compton process ($Q^{\rm c}$) and the bremsstrahlung process
($Q^{\rm eZ}$), as well as different values of the scaling dimension
$\dU$.}
\label{tab:redg}
\end{table}

One needs to observe that both bremsstrahlung and Compton processes play a role
in red giant environments. At typical horizontal-branch star densities $\rho
\approx 0.6 \times 10^4$  g/cm$^3$ and temperatures $T \approx 10^8 {\rm \ K} =
8.6$ keV, the bremsstrahlung process contributes roughly 10\% of the total
axion emission rate, while the Compton process accounts for 90\% of the rate
\cite{Raffelt:1994ry}. Then the bound $g_{aee} \lesim 2 \times 10^{-13}$
\cite{Raffelt:1999tx,Raffeltbook} translates into the limits in
Table~\ref{tab:comp}.

\paragraph{Constraints from stellar cooling on photon couplings.}

If unparticles only couple to photons, they would mainly contribute to star
cooling through the
process $\gamma+e \to e+{\cal U}$ via t-channel photon exchange (usually called
the Primakoff process), see Fig.~\ref{fg:feyn1}~(c). In the limit $\omega \ll
\me$ the cross-sections for this kind of process are
\begin{align}
\sigma^{\rm p}_{\gamma\gamma} &= 
c_{\gamma\gamma}^2/c_{\tilde{\gamma}\tilde{\gamma}}^2 \;
\sigma^{\rm p}_{\tilde{\gamma}\tilde{\gamma}} \\
&= 
  \frac{2\pi^2\alpha \, c_{\gamma\gamma}^2}{\kappa^2} \; 
  \begin{aligned}[t]
  &\biggl [ \frac{{}_3 F_2 (1,\tfrac{1+\dU}{2},1+\tfrac{\dU}{2};\tfrac{1}{2}+\dU,
    1+\dU; 4\omega^2/\kappa^2)}{\Gamma(2\dU)} \\
  &-  \frac{\dU \; {}_4 F_3 (1, \tfrac{1+\dU}{2}, 1+\tfrac{\dU}{2}, 1+\dU;
    \dU, \tfrac{1}{2}+\dU, 2+\dU; 4\omega^2/\kappa^2)}{(1+\dU)\Gamma(2\dU)}
  \biggr ]
  \left[\frac{\omega}{2\pi\MZ}\right]^{2\dU},
  \end{aligned} \displaybreak[0]
\end{align}
where $\kappa$ is the inverse Debye-H\"uckel radius, which accounts for the
screening of the Coulomb potential of the electron in a free stellar plasma
\cite{DH,Salpeter:1954nc,Raffelt:1985nk}. ${}_p F_q$ are generalized
hypergeometric functions.

In previous studies, limits have been derived for the coupling of axions to
photons. By comparing the unparticle production cross-section to the
cross-section for $\gamma+e \to e+a$,
\begin{equation}
\sigma^{\rm p}_{\rm a} = \frac{\alpha \, g_{a\gamma\gamma}^2}{8} \;
 \biggl [ 1 - \biggl ( 1+ \frac{\kappa^2}{4\omega^2} \biggr ) 
 \log \biggl ( 1+ \frac{4\omega^2}{\kappa^2} \biggr ) \biggr ],
\end{equation}
these limits can be translated to corresponding limits for the unparticle
couplings. Using $T \approx 10^8 {\rm \ K} = 8.6$ keV, $\kappa^2 = 7.5 \times
10^{-8}$ GeV$^2$
and $g_{a\gamma\gamma}\me < 5.5 \times 10^{-14}$ \cite{Raffelt:1985nk},
we find the limits in Table~\ref{tab:compg}.
Note that the dependence of the results on $\kappa$ is very mild; changing
$\kappa^2$ by an order of magnitude changes the limits in Table~\ref{tab:compg}
only by up to 20\%. Therefore these results should be reliable even
without a detailed numerical code for stellar evolution.


\subsubsection{Constraints from SN 1987A}
\label{sc:sn}

Unparticle emission would also influence supernova cooling. This has been
analyzed for vector unparticles in
Ref.~\cite{Davoudiasl:2007jr,Hannestad:2007ys}. Here the analysis in
Ref.~\cite{Hannestad:2007ys} is extended to derive limits for other types in
unparticle couplings, as in eq.~\eqref{eq:Uff}.

The observation of the length of the neutrino burst of the supernova SN 1987A
puts a strong constraint on the allowed energy loss rate due to unknown very
weakly interacting (un)particles \cite{Raffelt:1990yz},
\begin{equation}
Q_{\rm X} \lesim 3 \times 10^{33} {\rm\ erg\ cm}^{-3} {\rm\ s}^{-1}.
\end{equation}
Several processes can contribute to unparticle emission from the supernova core.
The dominant effect comes from neutron bremsstrahlung, $n+n \to n+n+{\cal U}$,
while proton bremsstrahlung is less important since
the proton density in supernova cores is smaller than the neutron density.
In principle bremsstrahlung processes with electrons, $e+n \to
e+n+{\cal U}$ and $e+e \to e+e+{\cal U}$, can be important due to collinear
enhancement. However, the collinear phase space region is suppressed due to
strong Coulomb screening effects in the dense core plasma, see e.g.
Ref.~\cite{Raffelt:1985nk}.

Since the supernova core temperature $T \approx 30$ MeV is much smaller than
the neutron mass, the neutron bremsstrahlung process factorizes into a "hard"
$nn$ collision process and "soft" unparticle radiation from one of the external
neutrons. Here one can distinguish between the case when the bremsstrahlung
coupling is insensitive to the nucleon spin (vector and scalar couplings of the
unparticles to the quarks) \cite{Hanhart:2000er} and when the bremsstrahlung
emission couples to the nucleon spin (axial-vector and pseudo-scalar couplings) 
\cite{Hanhart:2000ae}.

For vector and scalar unparticle-quark interactions,
one finds in the non-relativistic limit
\begin{align}
Q^{\rm nn}_{\rm {\cal U},V} &= 
\frac{C \, c_{\rm V}^2 \, m_{\rm n} \, \beta_i^7}{32\pi^{3/2}}\,
n_{\rm n}^2 \sigma^{\rm nn}_0
\frac{39+1073 \dU + 228 \dU^2 + 60 \dU^3}{(2\dU-1)(1+2\dU)
 (3+2\dU)(5+2\dU)\Gamma(2\dU+5/2)}
\left[\frac{m_{\rm n} \beta_i^2}{2\pi\MZ}\right]^{2\dU-2} ,
\label{eq:epsnnV}
 \displaybreak[0]\\
Q^{\rm nn}_{\rm {\cal U},S1} &= 
\frac{C \, c_{\rm S1}^2 \, m_{\rm n} \, \beta_i^3}{\pi^{3/2}}\,
n_{\rm n}^2 \sigma^{\rm nn}_0
\frac{2(21-8 \dU + 55 \dU^2 + 31 \dU^3 + 6 \dU^4)}{(2\dU-1)
 (1+2\dU)(3+2\dU)(5+2\dU)\Gamma(2\dU+5/2)}
\left[\frac{m_{\rm n} \beta_i^2}{2\pi\MZ}\right]^{2\dU} , 
\end{align}
where $\beta_i$ is the incident neutron velocity and $\sigma^{\rm nn}_0 \sim 25
\times 10^{-27}$ cm$^2$ is the typical $nn$ scattering cross section at the
given energy \cite{Hanhart:2000er}. $n_{\rm n} \approx 3 \times 10^{14} {\rm g}
{\rm cm}^{-3}$ denotes the neutron density.

Convolution with the Maxwellian thermal distribution gives
\begin{align}
Q^{\rm nn}_{\rm {\cal U},V}(T) &= 
\frac{C \, c_{\rm V}^2 \, T^{7/2}}{32\sqrt{2}\pi^2 \, m_{\rm n}^{5/2}}\,
n_{\rm n}^2 \sigma^{\rm nn}_0
\frac{(39+1073 \dU + 228 \dU^2 + 60 \dU^3)\Gamma(3+2\dU)}{(2\dU-1)
 (1+2\dU)(3+2\dU)(5+2\dU)\Gamma(2\dU+5/2)}
\left[\frac{T}{2\pi\MZ}\right]^{2\dU-2} , 
 \displaybreak[0]\\
Q^{\rm nn}_{\rm {\cal U},S1}(T) &= 
\frac{C \, c_{\rm S1}^2 \, T^{3/2}}{\sqrt{2}\pi^2 \, m_{\rm n}^{1/2}}\,
n_{\rm n}^2 \sigma^{\rm nn}_0
\frac{2(21-8 \dU + 55 \dU^2 + 31 \dU^3 + 6 \dU^4)\Gamma(3+2\dU)}{(2\dU-1)
 (1+2\dU)(3+2\dU)(5+2\dU)\Gamma(2\dU+5/2)}
\left[\frac{T}{2\pi\MZ}\right]^{2\dU} , 
\end{align}
Our result for $Q^{\rm nn}_{\rm {\cal U},V}(T)$ has the same dimensional
dependence as in Ref.~\cite{Hannestad:2007ys}, but we are able to identify an
additional numerical prefactor between 0.004 and 0.0014, depending on $\dU$.
Thus we arrive at somewhat weaker bounds for the unparticle interactions. In
addition we obtain bounds for scalar interaction between unparticles and
Standard Model fermions.
Assuming $C \ge 1$, the bounds in Table~\ref{tab:comp} are obtained.

For the emission of axial-vector and pseduo-scalar unparticles, the matrix
elements factorize in a similar way into the on-shell $nn$ collision process
and soft radiation from one of the external legs. Since the axial-vector and
pseduo-scalar unparticle emission couples to the spins of the nucleons, one
needs to take into account the spin dependence of the $nn$ transition, which is
given by the dynamical spin structure function
\cite{Hanhart:2000ae,Raffelt:1993ix}. Following
Ref.~\cite{Hanhart:2000ae,Raffelt:2006cw}, we obtain
\begin{align}
Q^{\rm nn}_{\rm {\cal U},A}(T) &=
\frac{C \, c_{\rm A}^2 \, T^2 \, n_{\rm n}}{4\pi^2 \, \Gamma(2\dU)}\,
\left[\frac{T}{2\pi\MZ}\right]^{2\dU-2} 
\int_0^\infty {\rm d}x \; x^{2 \dU} e^{-x} 
\frac{\Gamma_\sigma/T}{x^2 + (\Gamma_\sigma/2T)^2} \, , 
 \displaybreak[0]\\ 
Q^{\rm nn}_{\rm {\cal U},P1}(T) &=
\frac{3 \, C \, c_{\rm P1}^2 \, T^2 \, n_{\rm n}}{4(1+2\dU)\Gamma(2\dU)}
\left[\frac{T}{2\pi\MZ}\right]^{2\dU} 
\int_0^\infty {\rm d}x \; x^{2 \dU +2} e^{-x} 
\frac{\Gamma_\sigma/T}{x^2 + (\Gamma_\sigma/2T)^2} \, , 
 \displaybreak[0]\\
Q^{\rm nn}_{\rm {\cal U},P2}(T) &=
\frac{3 \, C \, c_{\rm P2}^2 \, T^2 \, n_{\rm n}}{(1+2\dU)\Gamma(2\dU)}
\left[\frac{T}{2\pi\MZ}\right]^{2\dU} 
\int_0^\infty {\rm d}x \; x^{2 \dU +2} e^{-x} 
\frac{\Gamma_\sigma/T}{x^2 + (\Gamma_\sigma/2T)^2} \, , 
\end{align}
where $\Gamma_\sigma$ is the spin fluctuation rate. Using a one-pion exchange
model for the nucleon scattering kernel,  one obtains the estimte
$\Gamma_\sigma \approx 450 {\rm \ MeV}$ for the typical temperature and density
inside the supernova core \cite{Raffelt:1993ix}. A more robust evaluation based
on experimental nucleon scattering data \cite{Hanhart:2000ae} finds a smaller
value for the spin structure function, which can be parametrized by using 
$
\Gamma_\sigma \approx 100 {\rm \ MeV}
$.
Taking this value and 
$C \ge 1$ as before, the bounds in Table~\ref{tab:comp} are derived.


\subsubsection{Comparison to reach of collider experiments and conclusions}
\label{sc:coll}

In Tables~\ref{tab:comp} and \ref{tab:compg} we summarize our limits on
unparticle couplings derived from astrophysical constraints. The bounds
correspond to the 90\% CL experimental error of the astrophysical
observations\footnote{In the derivation of the stellar energy loss constraints,
large systematic uncertainties could arise in the calculation of nuclear
interactions and stellar evolution. Since these errors are difficult to quantify
they have not been taken into account here.}, for the case that only one of the
unparticle couplings $c_{\rm X}$ is non-zero at a time. For comparison we also
show earlier results for limits from current (LEP, Tevatron) and future
colliders (LHC, ILC). To get an estimate of the possible reach of a future
international linear collider (ILC), we have assumed that it can  perform the
same kind of measurements as LEP, but with a 1000 times higher luminosity. Of
course, only  a proper analysis can go beyond this order-of-magnitude assessment
of the sensitivity of ILC. The blanks in the table indicate that no results are
available from the literature for the given interaction. Some of the processes
are not sensitive to a certain coupling, as denoted by a bar in the table.

It can be seen that the constraints for astrophysics are generally considerably
stronger than those from colliders. The strongest bounds are for vector/axial
couplings. For small $\dU$  limits on a 5th force are by far the dominant
constraints; however for $\dU$ tending towards two all constraints become
similarly important; here star cooling provides the strongest bound.  For scalar
and pseudoscalar couplings the bounds are generally weaker, which is mainly due
to the higher dimensionality of the interaction operators. For $\dU =1$,
the unparticle scaling behavior corresponds to a regular massless particle,
so that our limits also apply to any model which includes a new massless scalar or
vector particle (see also Ref.~\cite{Dobrescu:2004wz}).

For the unparticle-photon couplings, our bounds from star cooling are much
stronger than the limits from supernova cooling, taken from
Ref.~\cite{Das:2007nu}. These couplings could also be constrained by the process
$e^+e^- \to \gamma + {\cal U}$ at LEP and ILC, but this has not been analyzed so
far.

This analysis is restricted to the leading  CP-conserving and flavor-diagonal
unparticle interactions. The astrophysical constraints are not sensitive to
operators that involve flavor changing neutral currents, which can be tested in
precision experiments at low energies, such as heavy-flavor mixing and decays
\cite{Luo:2007bq,Chen:2007vv,Li:2007by,Chen:2007je,Aliev:2007gr,Mohanta:2007ad,Lenz:2007nj},
as well as to operators that only couple to third-generation fermions
\cite{Choudhury:2007cq,Alan:2007ss}, heavy gauge bosons \cite{Greiner:2007hr} or
the Higgs boson \cite{Kikuchi:2007qd,Chen:2007zy,Delgado:2007dx}. Furthermore,
direct CP-violation in the unparticle operators \cite{Zwicky:2007vv} can lead to
new effects, which cannot be tested in astrophysics.

\landscape

\renewcommand{\arraystretch}{1.2}
\begin{table}[p]
\begin{center}
\begin{tabular}{|l||c|c|c|c||c|c|c|c|l}
\cline{1-9}
Coupling & \multicolumn{4}{c||}{$c_{\rm V}$}  & 
	   \multicolumn{4}{c|}{$c_{\rm A}$} \\
\cline{1-9}
$\dU$ & 1 & 4/3 & 5/3 & 2 & 1 & 4/3 & 5/3 & 2 \\
\cline{1-9}
5th force ("E\"otv\"os") & $7 \cdot 10^{-24}$ & $1.4 \cdot 10^{-15}$ &
			   $1.8 \cdot 10^{-10}$ & $2 \cdot 10^{-5}$ &
			   $4 \cdot 10^{-24}$ & $8 \cdot 10^{-16}$ &
			   $1 \cdot 10^{-10}$ & $1.1 \cdot 10^{-5}$ \\
Energy loss from stars & $5 \cdot 10^{-15}$ & $2.5 \cdot 10^{-12}$ &
			   $1 \cdot 10^{-9}$ & $3.5 \cdot 10^{-7}$ &
			   $6.3 \cdot 10^{-15}$ & $2 \cdot 10^{-12}$ &
			   $7.3 \cdot 10^{-10}$ & $3 \cdot 10^{-7}$\\
SN 1987A & $1 \cdot 10^{-9}$ & $3.5 \cdot 10^{-8}$ &
			   $1 \cdot 10^{-6}$ & $3 \cdot 10^{-5}$ &
			   $2 \cdot 10^{-11}$ & $5.5 \cdot 10^{-10}$ &
			   $1.5 \cdot 10^{-8}$ & $4.1 \cdot 10^{-7}$ \\
\cline{1-9}
LEP & 0.005 & 0.045 & 0.04 & 0.01 & 0.1 & 0.045 & 0.04 & 0.008 &
	\cite{Cheung:2007ap,Bander:2007nd} \\
Tevatron & & 0.4 & 0.05 & & & & & & 
	\cite{Rizzo:2007xr} \\	
ILC & $1.6 \cdot 10^{-4}$ & $1.4 \cdot 10^{-3}$ &
			   $1.3 \cdot 10^{-3}$ & $3.2 \cdot 10^{-4}$ &
			   $3.2 \cdot 10^{-3}$ & $1.4 \cdot 10^{-3}$ &
			   $1.3 \cdot 10^{-3}$ & $2.5 \cdot 10^{-4}$ \\
LHC & & 0.25 & 0.02 & & & & & & 
	\cite{Rizzo:2007xr} \\			   
Electroweak precision & 1 & 0.2 & 0.025 & & 1 & 0.15 & 0.01 & &
	\cite{Liao:2007bx}\\
Quarkonia & & 0.01 & 0.1 & 0.45 & & & & & 
	\cite{Chen:2007zy} \\
Positronium & & 0.25 & & & & $2 \cdot 10^{-13}$ & $2 \cdot 10^{-8}$ & 0.03 &
	\cite{Liao:2007ic} \\
\cline{1-9}
\multicolumn{10}{c}{} \\[1.5em]
\cline{1-9}
Coupling & \multicolumn{4}{c||}{$c_{\rm S1}$}  & 
	   \multicolumn{4}{c|}{$c_{\rm P1}$, $2c_{\rm P2}$} \\
\cline{1-9}
$\dU$ & 1 & 4/3 & 5/3 & 2 & 1 & 4/3 & 5/3 & 2 \\
\cline{1-9}
5th force ("E\"otv\"os") & $6.5 \cdot 10^{-22}$ & $1.2 \cdot 10^{-13}$ &
			   $1.6 \cdot 10^{-8}$ & $1.7 \cdot 10^{-3}$ &
			   --- & --- & --- & --- \\
Energy loss from stars & $1.3 \cdot 10^{-9}$ & $7 \cdot 10^{-7}$ &
			   $3 \cdot 10^{-4}$ & $0.13$ &
			   $4 \cdot 10^{-8}$ & $1.1 \cdot 10^{-5}$ &
			   $3.3 \cdot 10^{-3}$ & $1$\\
SN 1987A & $8 \cdot 10^{-8}$ & $2.4 \cdot 10^{-6}$ &
			   $6.6 \cdot 10^{-5}$ & $2 \cdot 10^{-3}$ &
			   $5.5 \cdot 10^{-8}$ & $1.3 \cdot 10^{-6}$ &
			   $3.5 \cdot 10^{-5}$ & $9 \cdot 10^{-4}$ \\
\cline{1-9}
LEP & $>1$ & $>1$ & $>1$ & $>1$ & $>1$ & $>1$ & $>1$ & $>1$ & 
	\cite{Bander:2007nd}\\
ILC & $>1$ & $>1$ & $>1$ & $>1$ & $>1$ & $>1$ & $>1$ & $>1$ & 
	\\
\cline{1-9}
\end{tabular}
\end{center}
\mycaption{Comparison of limits for unparticle-fermion couplings 
from astrophysical constraints and from present and future
collider experiments. The astrophysical bounds have been derived in this work,
while the collider bounds have been taken from the literature, as indicated by
the references in the right column.
Blank spaces are left where no results are available from the literature,
while the bars denote that no bound on the coupling can be determined.}
\label{tab:comp}
\end{table}

\endlandscape


\renewcommand{\arraystretch}{1.2}
\begin{table}[tb]
\begin{center}
\begin{tabular}{|l||c|c|c|c|l}
\cline{1-5}
Coupling & \multicolumn{4}{c|}{$c_{\gamma\gamma},
c_{\tilde{\gamma}\tilde{\gamma}}$}  \\
\cline{1-5}
$\dU$ & 1 & 4/3 & 5/3 & 2 \\
\cline{1-5}
Energy loss from stars & $5.5 \cdot 10^{-14}$ & $1.7 \cdot 10^{-11}$ &
			   $5.3 \cdot 10^{-9}$ & $1.7 \cdot 10^{-6}$ \\
SN 1987A & $9 \cdot 10^{-7}$ & $4 \cdot 10^{-6}$ &
			   $4 \cdot 10^{-5}$ & $8 \cdot 10^{-4}$ &
	\cite{Das:2007nu}\\
\cline{1-5}
\end{tabular}
\end{center}
\mycaption{Comparison of limits for unparticle-photon couplings  from
astrophysical constraints. The
bounds from star cooling have been derived in this work, while the supernova 
bounds have been taken from the literature, as indicated by the
reference in the right column.}
\label{tab:compg}
\end{table}

\subsubsection*{Note added}

Shortly before finishing this manuscript, we became aware of related work on 5th
force experiments \cite{Deshpande:2007mf} where similar, though weaker limits
were obtained, since these  authors  included only
results from Newtonian-law experiments at short but not at astronomical
distances.


\subsubsection*{Acknowledgments}

This work was supported by the Schweizerischer Nationalfonds. We thank Pedro 
Schwaller for discussions. We are grateful to the journal referee for careful
reading and interesting and helpful comments.


\subsubsection*{Appendix: Phase-space integrals}

In the following the relevant phase-space integrals for the unparticle emission
processes in this article are summarized.

For the Compton process $e(p) + \gamma(k) \to e(p') + {\cal U}(k')$ in the
non-relativistic limit, with the initial photon energy $k_0 = \omega \ll \me$ it
is useful to choose a reference frame where the electron in the initial state is
at rest. The phase space integration then yields
\begin{align}
\sigma^{\rm c}_{\cal U} &= \frac{A_\dU}{4\me\omega} 
\int \frac{{\rm d}^3 p'}{(2\pi)^3} \frac{1}{2\me} 
\int \frac{{\rm d}^4 k'}{(2\pi)^4} \, \theta(k'_0) \theta(k'^2)\, 
(k'^2)^{\dU-2} \, (2\pi)^4 \delta^{(4)}(k'+p'-k-p)\, |{\cal M}|^2 
\nonumber \\
&= \frac{A_\dU}{32\pi^2 \me^2 \omega} \int_0^1 {\rm d}\cos\theta_{p'}
\int_0^{2\omega \cos\theta_{p'}} {\rm d}p' \, p'^2 \,
(2\omega p' \cos\theta_{p'} - p'^2)^{\dU-2} \, |{\cal M}|^2,
\end{align}
where $|{\cal M}|^2$ is the squared and spin-averaged matrix element, and
$\theta_{p'}$ is the angle between the incident photon and the outgoing
electron. The following integrals appear:
\begin{align}
&\int_0^{2\omega \cos\theta_{p'}} {\rm d}p' \, p'^{\dU+n} \,
(2\omega \cos\theta_{p'} - p')^{\dU-2}
= \frac{\Gamma(\dU+1) \Gamma(\dU+n+1)}{\Gamma(2\dU+n)}
 (2\omega \cos\theta_{p'})^{2\dU+n-1},
\\
&\int_0^1 {\rm d}\cos\theta_{p'} \, (\cos\theta_{p'})^{2\dU+n}
= \frac{1}{2\dU+n+1}
\end{align}
with $n=1,2,3,\dots$

For bremsstrahlung $e(p) + Z(q) \to e(p') + Z(q') + {\cal U}(k')$  one finds in
the non-relativistic limit
\begin{align}
\sigma^{\rm eZ}_{\cal U}  &=
\frac{A_\dU}{4 \me m_{\rm z} \beta_i} 
\int \frac{{\rm d}^3 p'}{(2\pi)^3} \frac{1}{2\me}
\int \frac{{\rm d}^3 q'}{(2\pi)^3} \frac{1}{2m_{\rm z}}
\int \frac{{\rm d}^4 k'}{(2\pi)^4} \, \theta(k'_0) \theta(k'^2)\, 
(k'^2)^{\dU-2} \nonumber \\
& \hspace{16em} \times (2\pi)^4 \delta^{(4)}(k'+p'+q'-p-q)\, |{\cal M}|^2
\nonumber  \\
&= \frac{A_\dU\me}{64\pi^4 m_{\rm z}^2 \beta_i}
\int_0^{\beta_i} {\rm d}\beta_f \, \beta_f^2 
\int \frac{{\rm d}\Omega_{p'}}{4\pi}
\int \frac{{\rm d}\Omega_{k'}}{4\pi}
\int_0^{\me(\beta_i^2-\beta_f^2)/2} {\rm d}|\vec{k}'| \, |\vec{k}'|^2 
\nonumber \\
& \hspace{16em} \times (\tfrac{1}{4}\me^2(\beta_i^2-\beta_f^2)^2 -|\vec{k}'|^2)^{\dU-2}  
\; |{\cal M}|^2,
\end{align}
where $\beta_{i,f}$ are the velocity of the incoming and outgoing electron,
respectively,
$\beta_i = |\vec{p}|/p_0$, $\beta_f = |\vec{p}'|/p'_0$, and $m_{\rm z}$ is the
mass of the nucleus. 
After including the matrix element, the typical integrals are
\begin{align}
&\int_0^{\frac{\me}{2}(\beta_i^2-\beta_f^2)} {\rm d}|\vec{k}'| \, |\vec{k}'|^n
\, (\tfrac{1}{4}\me^2(\beta_i^2-\beta_f^2)^2 -|\vec{k}'|^2)^{\dU-2}
\label{eq:iez1}
\nonumber \\[-1ex]
& \hspace{9em} = \frac{\Gamma(\dU-1)\Gamma(\frac{1+n}{2})}{2\,\Gamma(\dU+\frac{n-1}{2})}
 (\tfrac{\me}{2} (\beta_i^2-\beta_f^2))^{2\dU+n-4}, 
\qquad n=1,2,3,\dots,\\
&\int \frac{{\rm d}\Omega_{k'}}{4\pi} [\vec{k}' \cdot (\vec{p}-\vec{p}')]^2
= \tfrac{1}{3} \, |\vec{k}'|^2 (\vec{p}-\vec{p}')^2, \\
&\int \frac{{\rm d}\Omega_{p'}}{4\pi} \frac{1}{(\vec{p}-\vec{p}')^2}
= \frac{1}{\me^2 \, \beta_i\beta_f} 
 \log \frac{\beta_i+\beta_f}{\beta_i-\beta_f}, \\
&\int \frac{{\rm d}\Omega_{p'}}{4\pi} \frac{1}{(\vec{p}-\vec{p}')^4}
= \frac{2}{\me^4(\beta_i^2-\beta_f^2)^2}, \\
&\int_0^{\beta_i} {\rm d}\beta_f \, \beta_f^n \, (\beta_i^2-\beta_f^2)^{2\dU}
= - \frac{\pi \, \Gamma(\frac{1+n}{2}) \, \csc(2 \pi \dU)}
	{2 \, \Gamma(-2\dU) \Gamma(2\dU+\frac{n+3}{2})} \,
 \beta_i^{4\dU+n+1},\qquad n=1,2,3,\dots, \label{eq:iez5}
\\
&\int_0^{\beta_i} {\rm d}\beta_f \, \beta_f \, (\beta_i^2-\beta_f^2)^{2\dU}
 \, \log \frac{\beta_i+\beta_f}{\beta_i-\beta_f}
= \frac{\pi^{3/2} \csc(2 \pi \dU)}{2(2\dU+1) \, \Gamma(-2\dU) 
  \Gamma(2\dU+3/2)} \, \beta_i^{4\dU+2}.
\end{align}
Most of the above integrals are valid only for $\dU\ge 1$.

For bremsstrahlung off a neutron pair, $n(p) + n(q) \to n(p') + n(q') + {\cal
U}(k')$, the situation is very similar to electron-nucleus-bremsstrahlung,
albeit there are some differences due to fact that this process can only be
evaluated by factorizing the strongly interacting $nn \to nn$ scattering.
In the center-of-mass frame
\begin{align}
\frac{\sigma^{\rm nn}_{\cal U}}{\sigma^{\rm nn}_0}  &=
\frac{
A_\dU 
\int {\rm d}^3 p'
\int {\rm d}^3 q'
\int {\rm d}^4 k' \, \theta(k'_0) \theta(k'^2)\, 
(k'^2)^{\dU-2} \delta^{(4)}(k'+p'+q'-p-q)\, |{\cal M}|^2
}{
\int {\rm d}^3 p'
\int {\rm d}^3 q' \, (2\pi)^4
\delta^{(4)}(p'+q'-p-q)\, |{\cal M}_0|^2
} \nonumber \\
&= \frac{A_\dU m_{\rm n}}{4\pi^3\beta_i}
\int_0^{\beta_i} {\rm d}\beta_f \, \beta_f^2 
\int \frac{{\rm d}\Omega_{p'}}{4\pi}
\int \frac{{\rm d}\Omega_{k'}}{4\pi}
\int_0^{m_{\rm n}(\beta_i^2-\beta_f^2)} {\rm d}|\vec{k}'| \, |\vec{k}'|^2 
\nonumber \\
& \hspace{16em} \times (m_{\rm n}^2(\beta_i^2-\beta_f^2)^2 -|\vec{k}'|^2)^{\dU-2}  
\; |{\cal M}|^2/|{\cal M}_0|^2,
\end{align}
where ${\cal M}_0$ is the spin-averaged squared matrix element for $nn$
scattering, which in the non-relativistic limit does not depend on the
kinematic variables of the external particles, so that the integration in the
denominator is trivial. As before, $\beta_{i,f}$ are the velocities of the
incoming and outgoing neutrons, respectively.

Besides the integrals eqs.~\eqref{eq:iez1} [with $\me/2 \to m_{\rm n}$] and
\eqref{eq:iez5} one needs the integrals
\begin{align}
&\int \frac{{\rm d}\Omega_{k'}}{4\pi} (\vec{k}' \cdot \vec{p})^{2n}
= \tfrac{1}{2n+1} |\vec{k}'|^{2n} |\vec{p}'|^{2n}, \qquad n=1,2,3,\dots, \\
&\int \frac{{\rm d}\Omega_{k'}}{4\pi} (\vec{k}' \cdot \vec{p})^2
 (\vec{k}' \cdot \vec{p}')^2
= \tfrac{1}{60} [(\vec{p}-\vec{p}')^4 + (\vec{p}+\vec{p}')^4]
- \frac{1}{6} \int \frac{{\rm d}\Omega_{k'}}{4\pi} 
	[(\vec{k}' \cdot \vec{p})^4 + (\vec{k}' \cdot \vec{p}')^4].
\end{align}
To arrive at the cross-section formulae in sections~\ref{sc:stellar} and
\ref{sc:sn}, relations
between $\Gamma$-functions have been used extensively in some cases.


\begin{thebibliography}{99}
\frenchspacing

\bibitem{Georgi:2007ek}
  H.~Georgi,
  Phys.\ Rev.\ Lett.\  {\bf 98}, 221601 (2007)
  [arXiv:hep-ph/0703260].

\bibitem{vanderBij:2006ne}
  J.~J.~van der Bij,
  Phys.\ Lett.\  B {\bf 636}, 56 (2006)
  [arXiv:hep-ph/0603082].

\bibitem{Davoudiasl:2007jr}
  H.~Davoudiasl,
  arXiv:0705.3636 [hep-ph].

\bibitem{Hannestad:2007ys}
  S.~Hannestad, G.~Raffelt and Y.~Y.~Y.~Wong,
  arXiv:0708.1404 [hep-ph].

\bibitem{Chen:2007qr}
  S.~L.~Chen and X.~G.~He,
  arXiv:0705.3946 [hep-ph].

\bibitem{Okun:1969ey}
  L.~B.~Okun,
  Yad.\ Fiz.\  {\bf 10}, 358 (1969) [Sov.\ J.\ Nucl.\ Phys.\  {\bf 10}, 206
  (1969].

\bibitem{Fischbach:1992fa}
  E.~Fischbach and C.~Talmadge,
  Nature {\bf 356}, 207 (1992).

\bibitem{Eotvos:1922pb}
  R.~V.~E\"otv\"os, D.~Pek\'ar and E.~Fekete,
  Annalen Phys.\  {\bf 68}, 11 (1922).

\bibitem{Lee:1955vk}
  T.~D.~Lee and C.~N.~Yang,
  Phys.\ Rev.\  {\bf 98}, 1501 (1955).

\bibitem{Georgi:2007si}
  H.~Georgi,
  arXiv:0704.2457 [hep-ph].

\bibitem{Cheung:2007ue}
  K.~Cheung, W.~Y.~Keung and T.~C.~Yuan,
  arXiv:0704.2588 [hep-ph].

\bibitem{Adelberger:2006dh}
  E.~G.~Adelberger, B.~R.~Heckel, S.~Hoedl, C.~D.~Hoyle, D.~J.~Kapner and A.~Upadhye,
  Phys.\ Rev.\ Lett.\  {\bf 98}, 131104 (2007)
  [arXiv:hep-ph/0611223].

\bibitem{Kapner:2006si}
  D.~J.~Kapner, T.~S.~Cook, E.~G.~Adelberger, J.~H.~Gundlach, B.~R.~Heckel, C.~D.~Hoyle and H.~E.~Swanson,
  Phys.\ Rev.\ Lett.\  {\bf 98}, 021101 (2007)
  [arXiv:hep-ph/0611184].

\bibitem{Raffelt:1985nj}
  G.~G.~Raffelt,
  Phys.\ Lett.\  B {\bf 166}, 402 (1986).

\bibitem{Isern:1992}
  J.~Isern, M.~Hernanz and E.~Garc\'ia-Berro,
  Astrophys.\ J.\  {\bf 392}, L23 (1992).

\bibitem{Altherr:1993zd}
  T.~Altherr, E.~Petitgirard and T.~del Rio Gaztelurrutia,
  Astropart.\ Phys.\  {\bf 2}, 175 (1994)
  [arXiv:hep-ph/9310304].

\bibitem{Wang:1992kn}
  J.~Wang,
  Phys.\ Lett.\  B {\bf 291}, 97 (1992).

\bibitem{Raffelt:1990yz}
  G.~G.~Raffelt,
  Phys.\ Rept.\  {\bf 198}, 1 (1990);
  and references therein.

\bibitem{Raffelt:1999tx}
  G.~G.~Raffelt,
  Ann.\ Rev.\ Nucl.\ Part.\ Sci.\  {\bf 49}, 163 (1999)
  [arXiv:hep-ph/9903472].

\bibitem{Raffeltbook}
  G.~G.~Raffelt,
  {\it Stars as Laboratories for Fundamental Physics}, University of Chicago
  Press (1996).

\bibitem{Raffelt:1985nk}
  G.~G.~Raffelt,
  Phys.\ Rev.\  D {\bf 33}, 897 (1986).
  
\bibitem{Dearborn:1985gp}
  D.~S.~P.~Dearborn, D.~N.~Schramm and G.~Steigman,
  Phys.\ Rev.\ Lett.\  {\bf 56}, 26 (1986).

\bibitem{Raffelt:1994ry}
  G.~Raffelt and A.~Weiss,
  Phys.\ Rev.\  D {\bf 51}, 1495 (1995)
  [arXiv:hep-ph/9410205].
 
\bibitem{Krauss:1984gm}
  L.~M.~Krauss, J.~E.~Moody and F.~Wilczek,
  Phys.\ Lett.\  B {\bf 144}, 391 (1984).

\bibitem{DH}
P.~Debye and E.~H\"uckel, Phys.\ Z.\  {\bf 24}, 185 (1923).

\bibitem{Salpeter:1954nc}
  E.~E.~Salpeter,
  Austral.\ J.\ Phys.\  {\bf 7}, 373 (1954).

\bibitem{Hanhart:2000er}
  C.~Hanhart, D.~R.~Phillips, S.~Reddy and M.~J.~Savage,
  Nucl.\ Phys.\  B {\bf 595}, 335 (2001)
  [arXiv:nucl-th/0007016].

\bibitem{Hanhart:2000ae}
  C.~Hanhart, D.~R.~Phillips and S.~Reddy,
  Phys.\ Lett.\  B {\bf 499}, 9 (2001)
  [arXiv:astro-ph/0003445].

\bibitem{Raffelt:1993ix}
  G.~Raffelt and D.~Seckel,
  Phys.\ Rev.\  D {\bf 52}, 1780 (1995)
  [arXiv:astro-ph/9312019].

\bibitem{Raffelt:2006cw}
  G.~G.~Raffelt,
  arXiv:hep-ph/0611350.

\bibitem{Dobrescu:2004wz}
  B.~A.~Dobrescu,
  Phys.\ Rev.\ Lett.\  {\bf 94}, 151802 (2005)
  [arXiv:hep-ph/0411004].

\bibitem{Das:2007nu}
  P.~K.~Das,
  arXiv:0708.2812 [hep-ph].

\bibitem{Cheung:2007ap}
  K.~Cheung, W.~Y.~Keung and T.~C.~Yuan,
  arXiv:0706.3155 [hep-ph].

\bibitem{Bander:2007nd}
  M.~Bander, J.~L.~Feng, A.~Rajaraman and Y.~Shirman,
  arXiv:0706.2677 [hep-ph].

\bibitem{Rizzo:2007xr}
  T.~G.~Rizzo,
  arXiv:0706.3025 [hep-ph].

\bibitem{Liao:2007bx}
  Y.~Liao,
  arXiv:0705.0837 [hep-ph].

\bibitem{Chen:2007zy}
  S.~L.~Chen, X.~G.~He and H.~C.~Tsai,
  arXiv:0707.0187 [hep-ph].

\bibitem{Liao:2007ic}
  Y.~Liao and J.~Y.~Liu,
  arXiv:0706.1284 [hep-ph].

\bibitem{Luo:2007bq}
  M.~Luo and G.~Zhu,
  arXiv:0704.3532 [hep-ph].

\bibitem{Chen:2007vv}
 C.~H.~Chen and C.~Q.~Geng,
  arXiv:0705.0689 [hep-ph].

\bibitem{Li:2007by}
  X.~Q.~Li and Z.~T.~Wei,
  arXiv:0705.1821 [hep-ph].

\bibitem{Chen:2007je}
  C.~H.~Chen and C.~Q.~Geng,
  arXiv:0706.0850 [hep-ph].

\bibitem{Aliev:2007gr}
  T.~M.~Aliev, A.~S.~Cornell and N.~Gaur,
  JHEP {\bf 0707}, 072 (2007)
  [arXiv:0705.4542 [hep-ph]].

\bibitem{Mohanta:2007ad}
  R.~Mohanta and A.~K.~Giri,
  arXiv:0707.1234 [hep-ph].

\bibitem{Lenz:2007nj}
  A.~Lenz,
  arXiv:0707.1535 [hep-ph].

\bibitem{Choudhury:2007cq}
  D.~Choudhury and D.~K.~Ghosh,
  arXiv:0707.2074 [hep-ph].

\bibitem{Alan:2007ss}
  A.~T.~Alan and N.~K.~Pak,
  arXiv:0708.3802 [hep-ph].

\bibitem{Greiner:2007hr}
  N.~Greiner,
  arXiv:0705.3518 [hep-ph].

\bibitem{Kikuchi:2007qd}
  T.~Kikuchi and N.~Okada,
  arXiv:0707.0893 [hep-ph].

\bibitem{Delgado:2007dx}
  A.~Delgado, J.~R.~Espinosa and M.~Quiros,
  arXiv:0707.4309 [hep-ph].

\bibitem{Zwicky:2007vv}
  R.~Zwicky,
  arXiv:0707.0677 [hep-ph].

\bibitem{Deshpande:2007mf}
  N.~G.~Deshpande, S.~D.~H.~Hsu and J.~Jiang,
  arXiv:0708.2735 [hep-ph].

\end{thebibliography}
\end{document}